\documentclass{article}

\usepackage{CJKutf8}
\usepackage{PRIMEarxiv}
\usepackage{xcolor}
\usepackage[utf8]{inputenc} 
\usepackage[T1]{fontenc}    
\usepackage{hyperref}       
\usepackage{url}            
\usepackage{booktabs}       
\usepackage{amsfonts}       
\usepackage{nicefrac}       
\usepackage{microtype}      
\usepackage{lipsum}
\usepackage{fancyhdr}       
\usepackage{graphicx}       
\usepackage{authblk}
\usepackage{multirow}       
\graphicspath{{figs/}}      

\title{Leadsee-Precip: A Deep Learning Diagnostic Model \\ for Precipitation}

\author{ Weiwen Ji}
\author{ Jin Feng\thanks{Corresponding author: jfeng@ium.cn}}
\author{ Yueqi Liu}
\author{ Yulu Qiu}
\author{ Hua Gao}

\affil[]{Institute of Urban Meteorology, China Meteorological Administration, Beijing}

\begin{document}
\maketitle

\begin{abstract}

Recently, deep-learning weather forecasting models have surpassed traditional numerical models in terms of the accuracy of meteorological variables. However, there is considerable potential for improvements in precipitation forecasts, especially for heavy precipitation events. To address this deficiency, we propose Leadsee-Precip (\begin{CJK*}{UTF8}{gbsn}灵犀\end{CJK*}), a global deep learning model to generate precipitation from meteorological circulation fields. The model utilizes an information balance scheme to tackle the challenges of predicting heavy precipitation caused by the long-tail distribution of precipitation data. Additionally, more accurate satellite and radar-based precipitation retrievals are used as training targets. Compared to artificial intelligence global weather models, the heavy precipitation from Leadsee-Precip is more consistent with observations and shows competitive performance against global numerical weather prediction models. Leadsee-Precip can be integrated with any global circulation model to generate precipitation forecasts. But the deviations between the predicted and the ground-truth circulation fields may lead to a weakened precipitation forecast, which could potentially be mitigated by further fine-tuning based on the predicted circulation fields.

\end{abstract}

\section{Introduction}

Over the past two years, deep learning (DL) models have achieved remarkable advancements in weather prediction, particularly with the development of long-term datasets like the European Centre for Medium-Range Weather Forecasts (ECMWF) ERA5 \cite{hersbach2020era5}. FourCastNet \cite{pathak2022fourcastnet}, as the first high-resolution DL weather forecasting model, provides a short to medium-range global weather forecast at $0.25^\circ$ resolution, yet its accuracy still falls short of the performance of traditional numerical forecasting models, such as the Integrated Forecast Systems (IFS) of the ECMWF, which is widely recognized as the most accurate global weather forecast model \cite{haiden2024evaluation}. SwinRDM \cite{chen2023swinrdm} soon follows as a DL weather forecasting system to outperform ECMWF's high-resolution forecast (HRES) in 5-day forecasts with a spatial resolution of $0.25^\circ$. Pangu-Weather provides superior medium-range (7 days) forecasts at a resolution of $0.25^\circ$ than ECMWF HRES \cite{bi2023accurate}. GraphCast \cite{lam2022graphcast}, powered by graph neural networks, outperforms HRES in accuracy for 90\% of atmospheric variables in 10-day forecasts. FuXi \cite{chen2023fuxi} further extends the forecast range by an additional 15 days with a cascade structure. Besides, other DL models, such as FengWu \cite{chen2023fengwu} and AIFS \cite{lang2024aifs}, have introduced innovative advancements, delivering forecasts comparable to or even surpassing those of HRES. 
These DL models have shown remarkable prediction accuracy in circulation forecasts. However, their performance in predicting precipitation, particularly heavy precipitation events, remains unfavorable. Some DL models, such as Pangu-Weather and W-MAE \cite{man2023w}, exclude precipitation forecasting and focus on other atmospheric variables. Other models like FuXi and Graphcast do include precipitation predictions. However, ERA5 total precipitation used in these models is known to have biases \cite{lavers2022evaluation} and their published studies lack comprehensive evaluations of precipitation forecasts. Moreover, the distribution of precipitation across different magnitudes exhibits a long-tail distribution \cite{buda2018systematic}, which poses a significant challenge for DL models to accurately predict heavy rainfall. The long-tail data imbalance is commonly encountered in classification tasks within computer vision, which has led to many studies aiming to address this imbalance \cite{liu2019large,tang2020long,zhang2021distribution,zhang2022self}. However, precipitation forecast involves predicting continuous target values, shifting the challenge from classification to regression tasks. The long-tail imbalance in precipitation distribution often results in the DL models being biased towards more frequent moderate rainfall events, which in turn compromises their ability to accurately forecast the heavy precipitation that have rare samples \cite{wang2023customized,trebing2021smaatunetprecipitationnowcastingusing}. 
To Address the issue, we develop a new deep-learning precipitation diagnostic model, Leadsee-Precip, using an information balance (IB) scheme to address the long-tail distribution of precipitation data. Precipitation diagnoses are generated by input meteorological circulation fields with an encoder-decoder architecture \cite{badrinarayanan2017segnet}. The model extracts features from upper-air and surface variables separately and incorporates MogaNet \cite{li2023moganet}, a state-of-the-art backbone in computer vision tasks, as the bottleneck layer. Leadsee-Precip includes our self-developed IB scheme, assigning weights to different precipitations based on their frequency. This weighting scheme is applied during the backward propagation process when calculating the loss function and ensures that the loss function is more sensitive to rare and high-magnitude precipitations.
Additionally, we adopted a satellite-derived precipitation dataset covering the globe (from 60 degrees south to 60 degrees north) to replace the total precipitation in ERA5 as the target variable. Our model achieved a Threat Score (TS) of 0.185 and a Fraction Skill Score (FSS) of 0.570 for 6-hour accumulated diagnostic precipitation exceeding 25 mm on the test dataset. When validated using data from weather stations in China, the diagnostic precipitation combined with ERA5 circulation variables showed excellent Threat Scores (TS) in heavy precipitation events (intensity exceeding 25 $mm$ 6$h^{-1}$). Additionally, Leadsee-Precip can be integrated with the forecasts of any circulation model, however, this might result in a decline in its performance.
This is probably due to the differences between the circulation fields predicted by DL models and those from reanalysis datasets.
\section{Related Work}

FourCastNet \cite{pathak2022fourcastnet} trains a secondary precipitation diagnostic model to denote accumulated total precipitation forecast. GraphCast conducts the 6-hour precipitation forecast but does not claim performance due to data quality issues and a lack of fair baseline evaluation. It uses the Mean Squared Error (MSE) loss for precipitation training and treats precipitation like other variables. In MetNet-3 \cite{andrychowicz2023deep}, a regional deep learning weather forecast model using radar and satellite data as input, precipitation is converted from continuous values into a discrete distribution of 512 bins, transforming the task of precipitation forecasting into a classification problem. 

\section{Methods and Data}

\subsection{Model Architecture}
\begin{figure}
  \centering
  \includegraphics[width=13 cm]{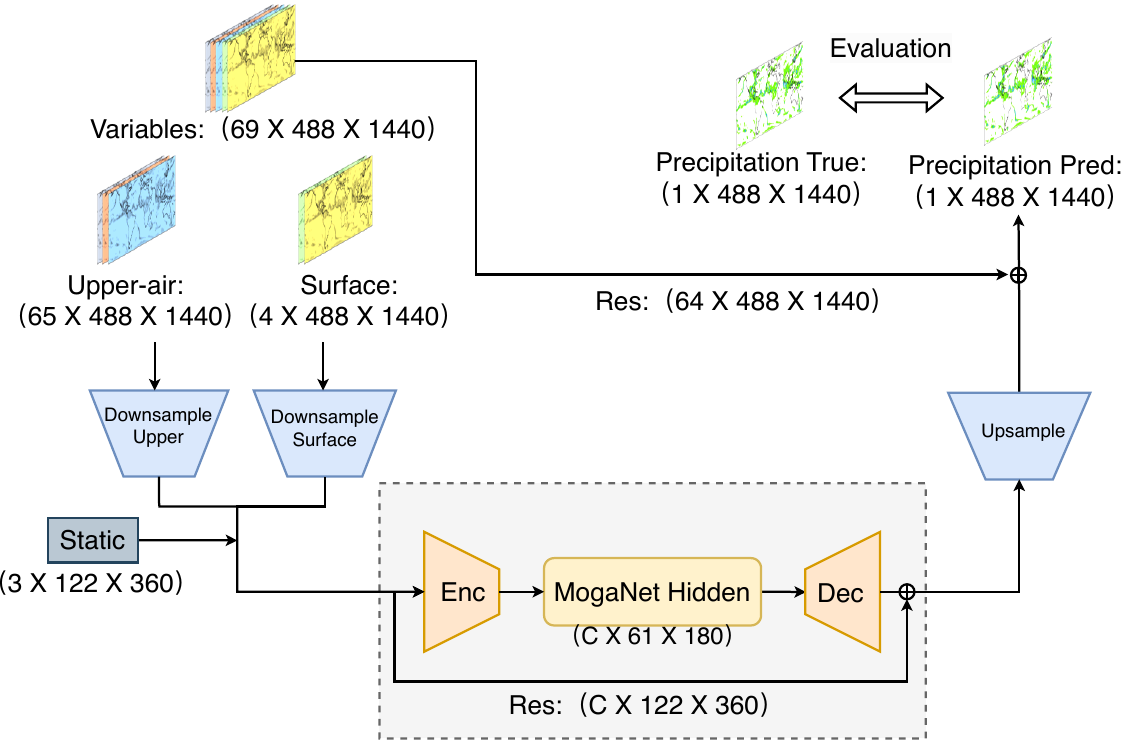}
  \caption{Structure of Leadsee-Precip. The model consists of feature extraction, hidden translator, and precipitation upsampling.}
  \label{fig:model_structure}
\end{figure}

Leadsee-Precip has an encoder-decoder structure as illustrated in Fig. \ref{fig:model_structure}. It consists of three parts: feature extraction, hidden translator, and precipitation upsampling. The feature extraction part primarily serves to derive features from circulation variables and reduce spatial resolution. To facilitate the model coupling, the input circulation variables consist of 69 channels, including 5 upper-air variables, each with 13 levels, and 4 surface variables, as the popular deep learning global atmospheric circulation prediction models. The upper-air variables are processed using 3D ConvNets for feature extraction and downsampling, allowing the model to capture the correlations across different variables and altitude levels. The surface variables use 2D ConvNets only. We set zonal circular padding in downsampling processes to ensure proper handling of the boundary conditions. After downsampling, the spatial resolution of upper-air and surface variables reduces to one-fourth. The downsampled upper-air and surface variables then concatenate together with an additional static layer, which includes three channels and has weights automatically learned by the model. A shortcut connection with 64 channels is established between the initial variables and the precipitation upsampling part to improve the accuracy of predictions.

The hidden translator part uses an encoder-decoder structure as well. The spatial resolution is reduced by half after the Enc module to save GPU memory usage while the number of channels remains unchanged. The MogaNet Hidden module contains 16 MogaNet blocks, which outperforms many of the currently leading Transformer-based network structures. The Dec module doubles the spatial resolution and adds the shortcut connection from the feature extraction part.

The precipitation upsampling part increases the spatial resolution of forecast results back to 0.25 degrees. A shortcut connection from the initial variables is added to this process to better reconstruct precipitation prediction. 

\begin{figure}
  \centering
  \includegraphics[width=11 cm]{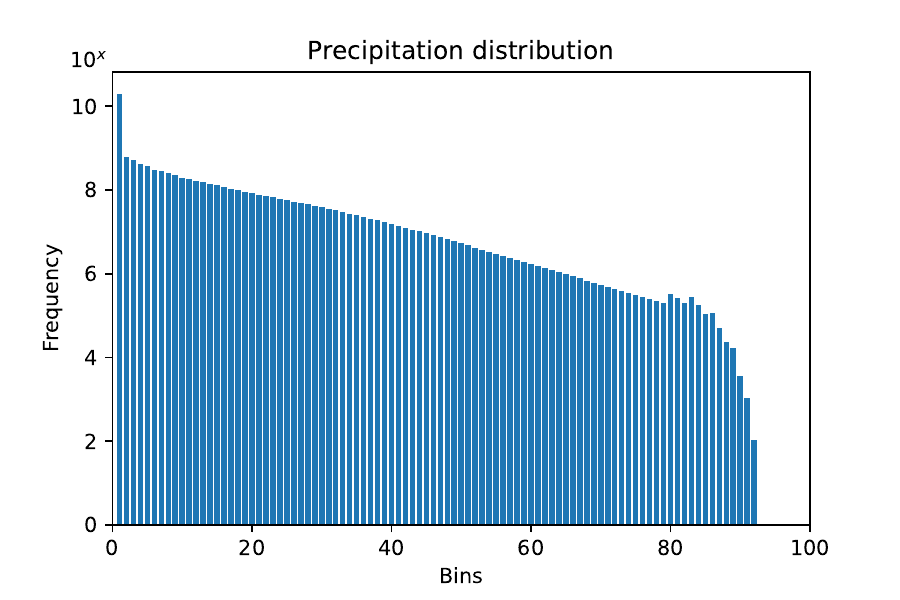}
  \caption{Distribution of precipitation. The precipitation data is categorized into 92 bins, with the precipitation intensity gradually increasing from zero to the maximum of 400 $mm$ $6h^{-1}$.}
  \label{fig:precip_dist}
\end{figure}

\subsection{Information Balance Scheme} \label{IB is designed and implemented by J.F.}
The long-tail distribution of precipitation data presents a challenge for DL models when directly trained using RMSE, as these models rarely sample the infrequent extreme precipitation events. This leads to biased forecasts, especially for heavy rainfall. The Google team introduced a method using the logit function to adjust the loss function (called logit adjustment loss), specifically aimed at addressing the class imbalance issue \cite{menon2020long}. Inspired by this approach, we develop an IB scheme specifically designed to address the regression challenges posed by long-tail data in precipitation forecasting. The IB scheme operates on the principle that no precipitation or light precipitation, due to their high probability of occurrence (low uncertainty), actually provides less information content. Therefore, the information content demonstrated using a logit form $-\log P(y_i)$ corresponding to the $i$-th precipitation sample can naturally serve as the weight for precipitation error, balancing the differences in sample sizes across different precipitation bins. The loss function of the model aims to minimize the total "information content-weighted error" across all precipitation samples. Specifically, The IB scheme uses the following formula to calculate the weight (\textit{$W_i$}):

\begin{equation}
    W_i = \frac{\left[-\log P\left(y_{i}\right)\right]^{\tau}}{\sum_{i}\left[-\log P\left(y_{i}\right)\right]^{\tau}}
\end{equation}

where the probability (\textit{$P(y_i)$}) statistically derived from the corresponding precipitation bin. We divide all precipitation grid point data into 92 bins by magnitude, as shown in Fig. \ref{fig:precip_dist}. This figure demonstrates that as precipitation magnitude increases, the frequency decreases, while the range of each bin gradually widens to achieve a more stable decline in frequency. Notice that the first bin, representing the range of 0 to 0.1, has a particularly high frequency, as most locations have a value of zero in precipitation data. 


We also introduce the hyper-parameter $\tau$, the temperature coefficient, following the logit adjustment method \cite{menon2020long}. The optimal value of this parameter is related to the target precipitation data and may vary across different studies. This effective weight is multiplied by the MSE during loss calculation, giving higher-magnitude precipitation a greater influence on the loss. Here we set its value to 2 after ablation experiments.  

\begin{figure}
    \centering
    \includegraphics[width=15cm]{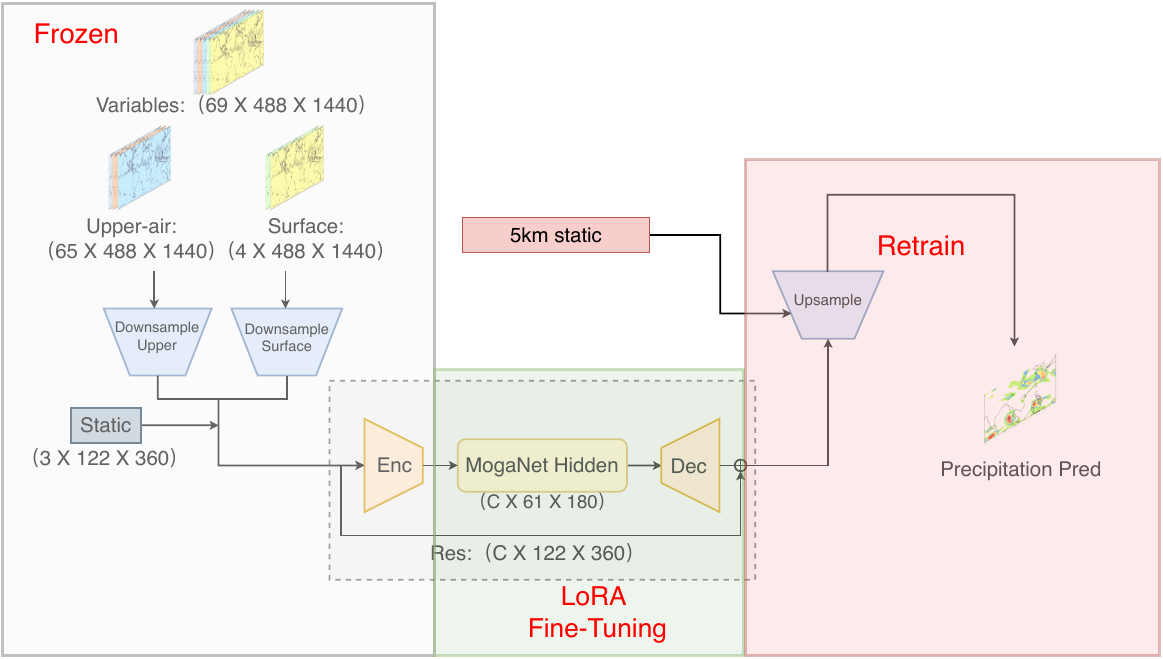}
    \caption{LoRA fine-tuning 5-km model structure. The feature extraction part is frozen, the MogaNet hidden part uses LoRA fine-tuning, and the upsampling part is retrained.}
    \label{fig:5km_lora}
\end{figure}

\subsection{Regional LoRA Fine-Tuning in China}
Based on the original Leadsee-Precip model, we applied LoRA fine-tuning and retrained it to achieve a 5 km resolution model for the China region (ranging from $15^\circ N$ to $55 N^\circ$ latitude and $70^\circ E$ to $140^\circ E$ longitude). 

Fig. \ref{fig:5km_lora} shows the fine-tuning structure. The input variables remain unchanged in the fine-tuning model, so we freeze the feature extraction part and the encoder module of the hidden translator part. LoRA fine-tuning is applied to the fully connected layers in the bottleneck and the convolutional layers in the decoder. The precipitation upsampling part is modified and retrained, with a static 5 km resolution topographic input added to enhance the details of the output prediction. To obtain the 5 km resolution forecast, we first crop the target region from the downsampled output of the hidden translator and then use PixelShuffle to upscale it to the target resolution. Note that the purpose of this study does not include downscaling. We use the simple PixelShuffle block is merely to match the resolution of the CoRA precipitation dataset over China (see Sect. \ref{CROA}).

\subsection{Evaluation Methods}
In this study, we utilize the Threat Score (TS, also known as the Critical Success Index, CSI) and the Fractions Skill Score (FSS) as key metrics to evaluate both the intensity and location of the precipitation. Traditional evaluation methods (such as root mean square error, RMSE) tend to favor smoother forecasts and are not able to account for the accuracy of the heavy precipitation \cite{mittermaier2013long}. For both TS and FSS, we select seven thresholds to evaluate the model's performance across varying precipitation intensities.

The TS method is used to evaluate binary precipitation forecasts and the formula is: 
\begin{equation}
    TS = \frac{Hits}{Hits+False\ Alarms+Missings},
\end{equation}
where Hits represent the total number of correct event forecasts, False Alarms are samples that which precipitation is incorrectly predicted, and Missings are precipitation events that are neglected. TS ranges from 0 to 1, with higher values indicating better predictions. 

The FSS method assesses the spatial accuracy of precipitation forecasts by comparing local observation and forecast in pixel windows (we use a window size of 7 pixels with average pooling). The formula of FSS is given by:
\begin{equation}
    FSS = 1- \frac{\sum{(P-T)^2}}{\sum{P^2}+\sum{T^2}},
\end{equation}
where P and T are the predicted and true fractions respectively at each grid point. The FSS value also ranges from 0 to 1, where values closer to 0 indicate a complete mismatch and a value of 1 represents a perfect forecast. 

\subsection{Datasets}

\subsubsection{ERA5}
In this work, we use circulation fields from the fifth generation of the ECMWF reanalysis dataset (ERA5). We select 5 upper-air atmospheric variables, each consisting of 13 pressure levels (1000, 925, 850, 700, 600, 500, 400, 300, 250, 200, 150, 100, and 50 hPa), and 4 surface variables at a spatial resolution of $0.25^\circ$ with a global grid size of 721 $\times$ 1440 pixels. The upper-air variables are geopotential (Z), specific humidity (Q), temperature (T), u-component of wind (U), and v-component of wind (V). The surface variables are 10-meter u-component of wind ($U_{10}$), 10-meter v-component of wind ($V_{10}$), 2-meter temperature ($T_{2m}$), and mean sea level pressure (MSLP). Due to the constraints in data download progress, we use a mixed temporal resolution for ERA5 data, using six-hour intervals from 1998 to 2012 and one-hour intervals from 2013 to 2022. Data from April to September 2022 is used for testing while the rest is allocated for training, as this model is primarily designed to capture extreme precipitation events in mid-latitude regions of the north hemisphere such as China.

\subsubsection{NOAA CMORPH}
The reference precipitation data used in this study is derived from the original National Oceanic and Atmospheric Administration Climate Prediction Center Morphing Technique (NOAA CMORPH) dataset \cite{joyce2004cmorph}, which provides global (60 degrees south - 60 degrees north) rain rate measurements at a 30-minute temporal and 8 km spatial resolution. We extract the rain rate values at each hour to represent hourly precipitation. After aggregating these values over six-hour intervals, we applied a max-pooling interpolation method to generate the final 6-hour accumulated target precipitation at a $0.25^\circ$ spatial resolution. The train-test split of NOAA CMORPH is the same as the ERA5 dataset.

\subsubsection{CROA} \label{CROA}
We use the high-resolution China Regional Observational Analysis dataset (CROA) from the China Meteorological Administration (CMA) for fine-tuning models. The data has a spatial resolution of 5 km, covering $15^\circ N$ to $55^\circ N$ and $70^\circ E$ to $140^\circ E$, with an hourly temporal resolution from 2013 to 2021. Each precipitation data point represents the accumulation over the previous 6 hours up to the current time, just like the CMORPH target precipitation above. This high-resolution dataset enhances the model's ability to improve finer-scale precipitation patterns in the China region. We utilize data from April to September of 2021 for the test and the rest for the train. 

\subsubsection{Weather Station Data} \label{Weather_Station_Data}
We use weather station precipitation data from about 10470 weather stations in the China \cite{feng2024enhancing} to conduct extra evaluations. The original data is stored in hourly precipitations and is processed to 6-hour cumulative ones for the test. When evaluating with weather station data, only the precipitation results at the station location are used.  It is important to note that we did not use the weather station dataset to train the model. Therefore, such evaluation based on “third-part” weather stations dataset is more objective but challenging.

\section{Results}
Here, we present the evaluation results of Leadsee-Precip on the global test set, followed by verification results using weather station data, and finally, the evaluation results with weather station data of the fine-tuning model. 

\subsection{Global Evaluation Using CMORPH Data}

\begin{figure}
    \centering
    \includegraphics[width=15cm]{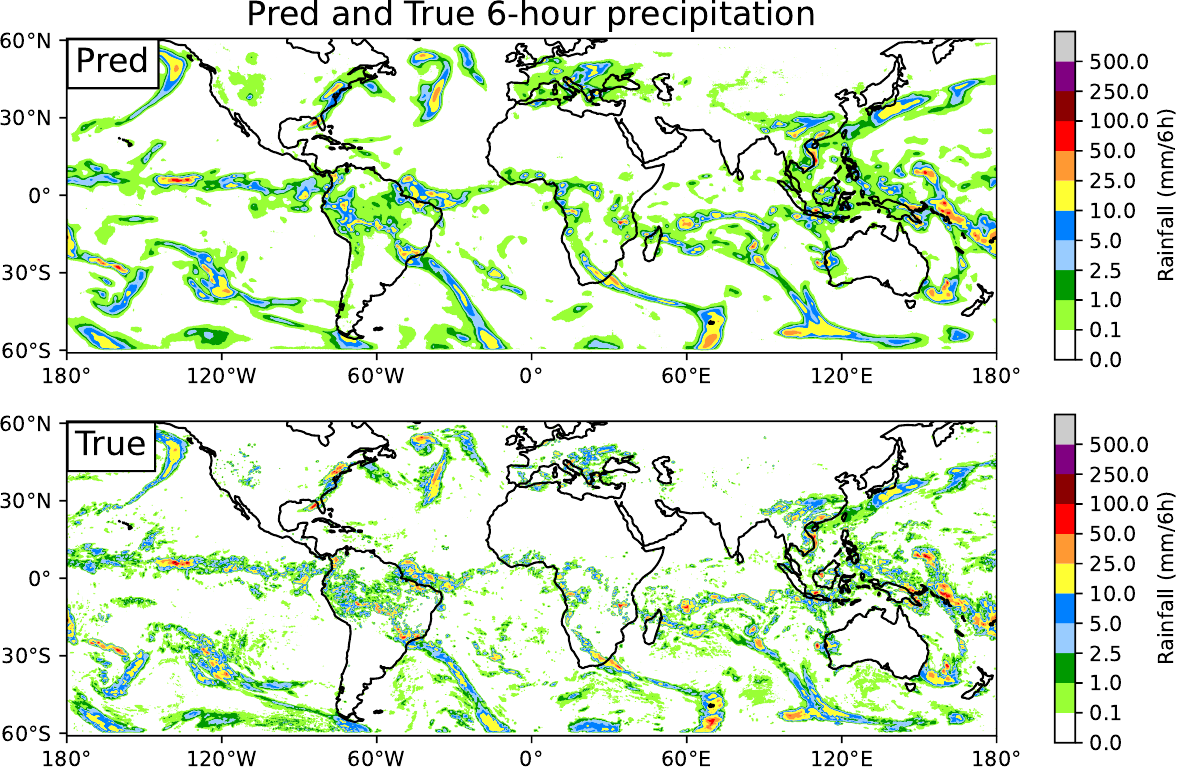}
    \caption{An illustrative example of a global 6-hour accumulated precipitation prediction generated by Leadsee-Precip and the ground truth of NOAA CMORPH. The calendar time-stamp of the figure was 00:00 UTC on April 1, 2022.}
    \label{fig:25km_Pred_True}
\end{figure}

\begin{table}
\setlength{\tabcolsep}{10pt} 
\renewcommand{\arraystretch}{1.5} 
    \centering
    \caption{Global evaluation metrics using CMORPH}
    \begin{tabular}{lccccccc}

        \toprule %
        Thresholds ($mm$ $6h^{-1}$) &0.1 & 1 & 5 & 10 & 25 & 50 & 100\\
        \midrule %
        TS &0.527 & 0.528 & 0.437 & 0.351 & 0.185 & 0.058 & 0.003 \\
        FSS &0.820 & 0.850 & 0.809 & 0.750 & 0.570 & 0.255 & 0.013 \\
        \bottomrule %
    \end{tabular}
    
    \label{tab:TS_FSS_25km}
\end{table}

Fig. \ref{fig:25km_Pred_True} shows the diagnosing skill of our model over the global range of $60^\circ S$ to $60^\circ N$. The precipitation in the upper panel is diagnosed by the model based on 69 variables from upper-air layers and surface data at a single time step while the lower panel illustrates the ground truth of the 6-hour accumulated precipitation of NOAA CMORPH. The diagnosed large rainfall (e.g., above the 25 $mm$ $6h^{-1}$ threshold) shows good consistency with the ground truth in both intensity and location. For instance, the model accurately reproduces several events of 6-hour accumulated rainfall exceeding 50 $mm$ in the eastern and western Pacific regions. For precipitation below 1 $mm$ $6h^{-1}$, the diagnostic results show a tendency for overestimation, with finer details appearing comparatively smoothed. 

Table \ref{tab:TS_FSS_25km} presents the evaluation metrics TS and FSS of the model on the test dataset. We selected seven 6-hour accumulated precipitation thresholds: 0.1, 1, 5, 10, 25, and 50 $mm$ $6h^{-1}$. The metrics generally show a decline with increasing precipitation levels while both TS and FSS slightly increase at the 0.1 and 1 $mm$ $6h^{-1}$ thresholds, suggesting the model tends to overestimate small rainfalls. However, the FSS of 25 $mm$ $6h^{-1}$ still exceeds 0.5, highlighting the model's ability to capture heavy precipitation events. 

\subsection{Regional Evaluation Using Weather Station Data}
\begin{figure}
    \centering
    \includegraphics[width=15 cm]{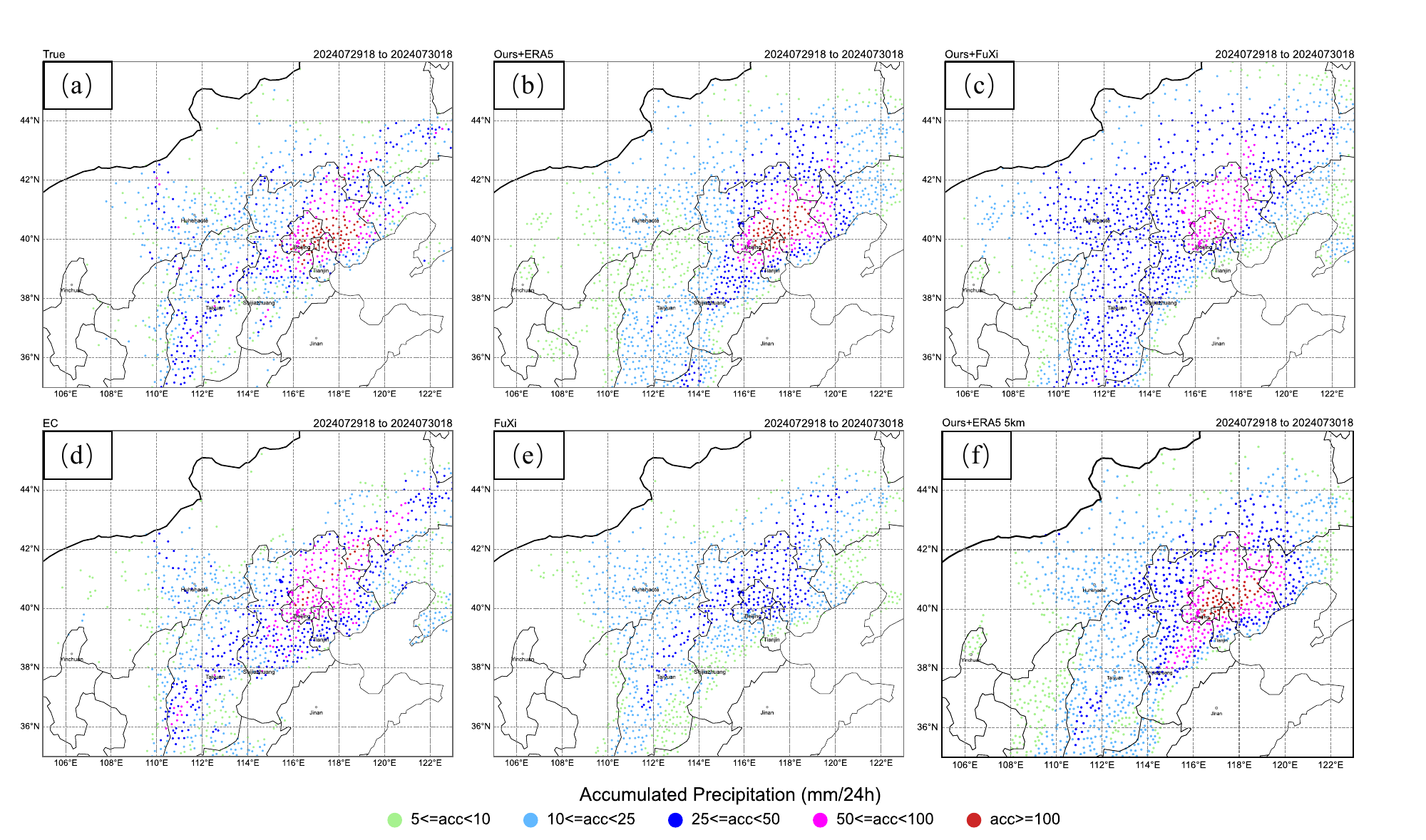}
    \caption{24-hour accumulated precipitation at weather station locations: observed and different model results. Panel (a) shows the ground truth of weather station data. Panel (b) shows the diagnosed precipitation of Leadsee. Panel (c) shows the Leadsee-Precip forecast driven by FuXi. Panel (d) shows the precipitation forecast of ECMWF HRES. Panel (e) shows the native precipitation forecast of FuXi. Panle (f) shows the fine-tuning Leadsee-Precip diagnosis. The accumulated precipitation covers the period from 18:00 UTC on July 29, 2024, to 18:00 UTC on July 30, 2024.}
    \label{fig:25km_jjj_models_site_eva}
\end{figure}

We use weather station data from the China region to evaluate the accuracy of Leadsee-Precip in four thresholds (5, 10, 25, and 50 $mm$). Table \ref{tab:25km_model_site_eva} presents the evaluation of four 6-hour precipitation diagnoses. Although the differences between the station data and the train-test data do pose challenges to the model's accuracy, Leadsee-Precip demonstrates slightly inferior TS scores above the 25 mm threshold compared with Table \ref{tab:TS_FSS_25km}. 
Fig. \ref{fig:25km_jjj_models_site_eva} presents a 24-hour accumulated precipitation event at weather station locations in the North China region across different models. The observed precipitation (Fig. \ref{fig:25km_jjj_models_site_eva} (a)) reveals that the heavy rainfall band (larger than 25 $mm$) extends in the northeast-southwest direction, with intense precipitation exceeding 50 $mm$ concentrated in the Beijing area and the northwestern part of Hebei province. Despite some contraction and deviation in the rainband exceeding 25 $mm$, the Leadsee-Precip diagnostic precipitation results (Fig. \ref{fig:25km_jjj_models_site_eva} (b)) align most closely with observed data for higher precipitation intensities, accurately capturing the location and strength of rainfall in the 50 $mm$ and 100 $mm$ thresholds. When driven by FuXi, Leadsee-Precip precipitation forecasts (Fig. \ref{fig:25km_jjj_models_site_eva} (c)) tend to underestimate precipitation amounts above 25 $mm$, and its ability to forecast heavy rainfall is slightly reduced. However, it still outperforms FuXi's native precipitation forecasts (Fig. \ref{fig:25km_jjj_models_site_eva} (e)), especially for rainfall exceeding 50 $mm$, with the maximum precipitation value in FuXi's native forecasts being even lower than 50 $mm$. ECMWF HRES precipitation forecasts (Fig. \ref{fig:25km_jjj_models_site_eva} (d)) demonstrate excellent accuracy in both the direction of the rainband and the location of heavy rainfall except for some misses of extreme precipitations (over 100 $mm$). In Fig. \ref{fig:25km_jjj_models_site_eva}, Leadsee+ERA5 and Leadsee+ERA5 5km uses circulation fields at the corresponding time, Leadsee+FuXi and FuXi utilize a lead time of 0, and EC uses a lead time of 6 hours as we can only access EC precipitation forecast starting from 00:00 and 12:00 UTC. 

\begin{table}
\setlength{\tabcolsep}{10pt} 
\renewcommand{\arraystretch}{1.5} 
    \centering
    \caption{Evaluation metrics using weather station data from July to August 2024 over China} 
    \begin{tabular}{lcccc}
    \toprule %
    Threshold ($mm$ $6h^{-1}$) & 5 & 10 & 25 & 50 \\
    \midrule %
    TS & 0.25 & 0.20 & 0.11 & 0.04 \\
    Bias & 1.59 & 1.33 & 0.68 & 0.17 \\
    \bottomrule %
    \end{tabular}
    
    \label{tab:25km_model_site_eva}
\end{table}

\subsection{Regional Evaluation of the Fine-Tuning Model}

Table \ref{tab:5kmLoRA_Eva} shows the evaluation results of the LoRA fine-tuning model, which has higher TS scores and lower bias in all thresholds than the model without fine-tuning. Fig. \ref{fig:25km_jjj_models_site_eva} (f) 
shows the fine-tuning Leadsee-Precip diagnostic results for the same accumulated precipitation event as in Fig. \ref{fig:25km_jjj_models_site_eva}. The fine-tuning precipitation diagnosis more accurately captures the location of the rainband exceeding 25 $mm$ $24 h^{-1}$  and the heavy precipitation areas over 100 $mm$ $24 h^{-1}$, compared to the Leadsee-Precip model without fine-tuning (Fig. \ref{fig:25km_jjj_models_site_eva} (b)), and is closer to the observed data (Fig. \ref{fig:25km_jjj_models_site_eva} (a)). 

\begin{table}
\setlength{\tabcolsep}{10pt} 
\renewcommand{\arraystretch}{1.5} 
    \centering
    \caption{Evaluation metrics of fine-tuning model using weather station data from July to August 2024 over China}
\begin{tabular}{lccccc}
    \toprule %
    Threshold ($mm$ $6h^{-1}$) & 5 & 10 & 25 & 50 \\
    \midrule %
    TS & 0.29 & 0.25 & 0.14 & 0.06 \\
    Bias & 1.49 & 1.16 & 0.52 & 0.16\\
    \bottomrule %
\end{tabular}
    
    \label{tab:5kmLoRA_Eva}
\end{table}

\begin{figure}
    \centering
    \includegraphics[width=15cm]{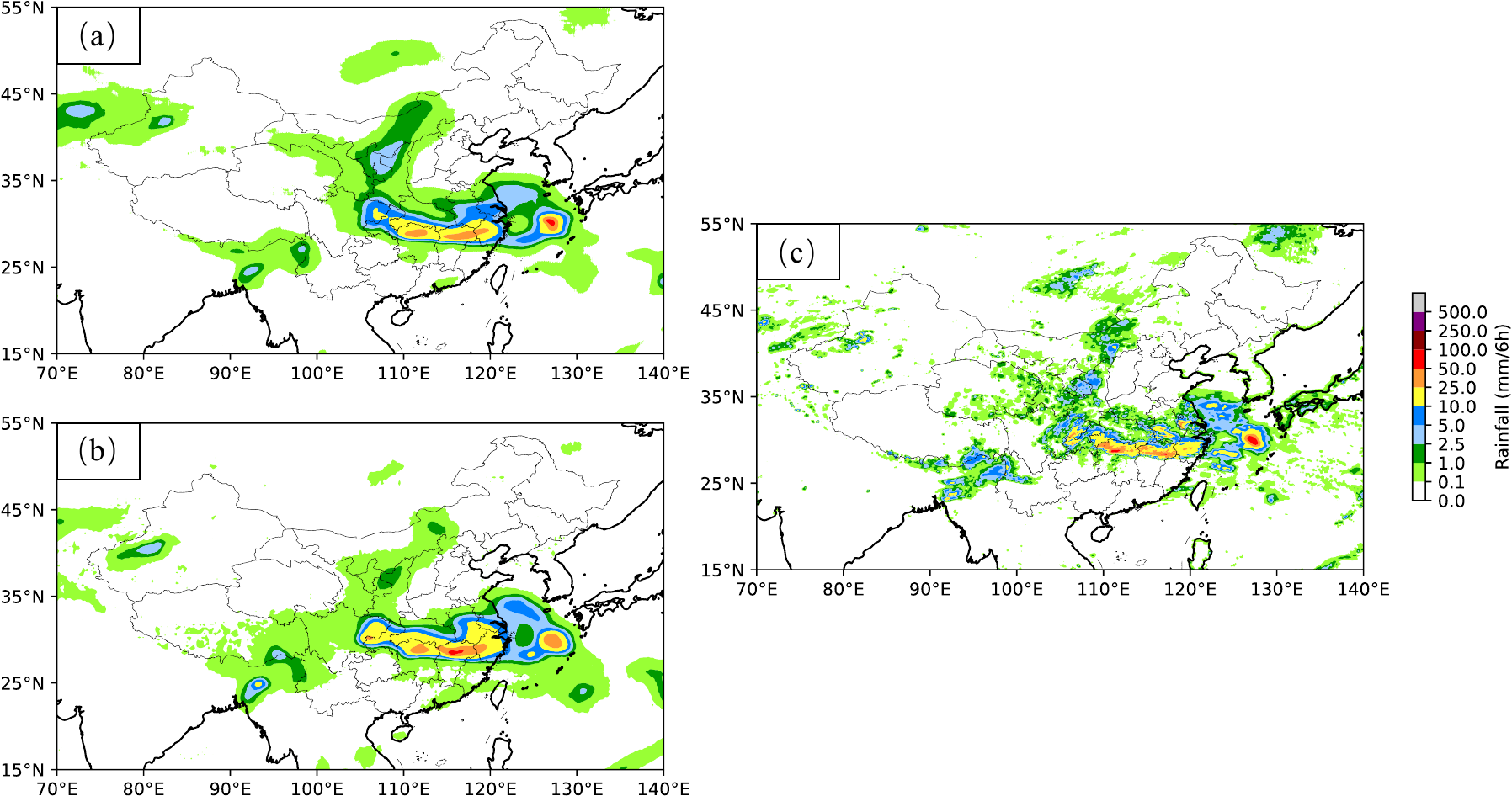}
    \caption{Comparison of 6-hour precipitation results between fine-tuning and initial Leadsee-Precip model. Panel (a) shows the precipitation from Leadsee-Precip with fine-tuning. Panel (b) shows the precipitation from the original Leadsee-Precip. Panel (c) shows the ground truth precipitation from the China regional observational analysis dataset. The 6-hour precipitation covers the period from 00:00 UTC on April 1, 2021, to 06:00 UTC on April 1, 2021.}
    \label{fig:5kmLoRA_25km_true}
\end{figure}

Fig. \ref{fig:5kmLoRA_25km_true} illustrates a 6-hour precipitation diagnosis comparison between the Leadsee-Precip model with and without fine-tuning. The precipitation diagnosis from the fine-tuning model (Fig. \ref{fig:5kmLoRA_25km_true} (a)) is more consistent with the ground truth (Fig. \ref{fig:5kmLoRA_25km_true} (c)) in terms of overall distribution and pattern. However, there is no significant improvement in the heavy precipitation forecast compared to the original results (Fig. \ref{fig:5kmLoRA_25km_true} (b)). Although the fine-tuned Leadsee-Precip model generates precipitation diagnosis at a 5 km resolution, the results reveal no significant improvement in real resolution compared to the original 25 km product, which indicates a need for further refinement in future work.

\section{Conclusion and Discussion}
In this work, we developed a precipitation diagnostic model, Leadsee-Precip, which employs an information balancing scheme and satellite precipitation data to improve the forecast accuracy in heavy precipitation. The fine-tuning Leadsee-Precip model using CROA dataset in China further enhances the precipitation patterns, and hence improves precipitation TS scores. This model can also be integrated with any circulation model to generate precipitation forecasts such as Fuxi. Combined with FuXi circulation fields, the TS scores of precipitation are less than the diagnostic one but can still catch the heavy rainfall events in North China.

It should be noted that changes in circulation fields may lead to additional errors in precipitation forecasting. For instance, when precipitation is generated using the FuXi circulation fields, the precipitation intensity is significantly weaker compared to that generated using reanalysis circulation fields. This discrepancy is likely attributed to the differences in energy at meso- and small-scales between the circulation fields predicted by the deep learning model and the reanalysis data. To further enhance the performance of precipitation forecasting, it is necessary to fine-tune Leadsee-Precip based on specific circulation model data, so that the precipitation generated by the model can better adapt to the circulation fields.

\section*{Acknowledgments}
This study is supported by the National Science Foundation of China (42275009), the Science and Technology Project of Beijing Meteorological Service (BMBKJ202401001), and the Special Project of Central Government Guiding Local Science and Technology Development in Beijing 2024 (Z241100001324014). We appreciate the National Meteorological Information Center of the China Meteorological Administration (NMIC/CMA) for CROA data of the China region. 


\bibliographystyle{unsrt}  

\begin{thebibliography}{10}

\bibitem{hersbach2020era5}
Hans Hersbach, Bill Bell, Paul Berrisford, Shoji Hirahara, Andr{\'a}s Hor{\'a}nyi, Joaqu{\'\i}n Mu{\~n}oz-Sabater, Julien Nicolas, Carole Peubey, Raluca Radu, Dinand Schepers, et~al.
\newblock The era5 global reanalysis.
\newblock {\em Quarterly Journal of the Royal Meteorological Society}, 146(730):1999--2049, 2020.

\bibitem{pathak2022fourcastnet}
Jaideep Pathak, Shashank Subramanian, Peter Harrington, Sanjeev Raja, Ashesh Chattopadhyay, Morteza Mardani, Thorsten Kurth, David Hall, Zongyi Li, Kamyar Azizzadenesheli, et~al.
\newblock Fourcastnet: A global data-driven high-resolution weather model using adaptive fourier neural operators.
\newblock {\em arXiv preprint arXiv:2202.11214}, 2022.

\bibitem{haiden2024evaluation}
Thomas Haiden, Martin Janousek, Frederic Vitart, Maliko Tanguy, Fernando Prates, and Matthieu Chevallier.
\newblock {\em Evaluation of ECMWF forecasts}.
\newblock European Centre for Medium Range Weather Forecasts Reading, UK, 2024.

\bibitem{chen2023swinrdm}
Lei Chen, Fei Du, Yuan Hu, Zhibin Wang, and Fan Wang.
\newblock Swinrdm: integrate swinrnn with diffusion model towards high-resolution and high-quality weather forecasting.
\newblock In {\em Proceedings of the AAAI Conference on Artificial Intelligence}, volume~37, pages 322--330, 2023.

\bibitem{bi2023accurate}
Kaifeng Bi, Lingxi Xie, Hengheng Zhang, Xin Chen, Xiaotao Gu, and Qi~Tian.
\newblock Accurate medium-range global weather forecasting with 3d neural networks.
\newblock {\em Nature}, 619(7970):533--538, 2023.

\bibitem{lam2022graphcast}
Remi Lam, Alvaro Sanchez-Gonzalez, Matthew Willson, Peter Wirnsberger, Meire Fortunato, Ferran Alet, Suman Ravuri, Timo Ewalds, Zach Eaton-Rosen, Weihua Hu, et~al.
\newblock Graphcast: Learning skillful medium-range global weather forecasting.
\newblock {\em arXiv preprint arXiv:2212.12794}, 2022.

\bibitem{chen2023fuxi}
Lei Chen, Xiaohui Zhong, Feng Zhang, Yuan Cheng, Yinghui Xu, Yuan Qi, and Hao Li.
\newblock Fuxi: A cascade machine learning forecasting system for 15-day global weather forecast.
\newblock {\em npj Climate and Atmospheric Science}, 6(1):190, 2023.

\bibitem{chen2023fengwu}
Kang Chen, Tao Han, Junchao Gong, Lei Bai, Fenghua Ling, Jing-Jia Luo, Xi~Chen, Leiming Ma, Tianning Zhang, Rui Su, et~al.
\newblock Fengwu: Pushing the skillful global medium-range weather forecast beyond 10 days lead.
\newblock {\em arXiv preprint arXiv:2304.02948}, 2023.

\bibitem{lang2024aifs}
Simon Lang, Mihai Alexe, Matthew Chantry, Jesper Dramsch, Florian Pinault, Baudouin Raoult, Mariana~CA Clare, Christian Lessig, Michael Maier-Gerber, Linus Magnusson, et~al.
\newblock Aifs-ecmwf's data-driven forecasting system.
\newblock {\em arXiv preprint arXiv:2406.01465}, 2024.

\bibitem{man2023w}
Xin Man, Chenghong Zhang, Jin Feng, Changyu Li, and Jie Shao.
\newblock W-mae: Pre-trained weather model with masked autoencoder for multi-variable weather forecasting.
\newblock {\em arXiv preprint arXiv:2304.08754}, 2023.

\bibitem{lavers2022evaluation}
David~A Lavers, Adrian Simmons, Freja Vamborg, and Mark~J Rodwell.
\newblock An evaluation of era5 precipitation for climate monitoring.
\newblock {\em Quarterly Journal of the Royal Meteorological Society}, 148(748):3152--3165, 2022.

\bibitem{buda2018systematic}
Mateusz Buda, Atsuto Maki, and Maciej~A Mazurowski.
\newblock A systematic study of the class imbalance problem in convolutional neural networks.
\newblock {\em Neural networks}, 106:249--259, 2018.

\bibitem{liu2019large}
Ziwei Liu, Zhongqi Miao, Xiaohang Zhan, Jiayun Wang, Boqing Gong, and Stella~X Yu.
\newblock Large-scale long-tailed recognition in an open world.
\newblock In {\em Proceedings of the IEEE/CVF conference on computer vision and pattern recognition}, pages 2537--2546, 2019.

\bibitem{tang2020long}
Kaihua Tang, Jianqiang Huang, and Hanwang Zhang.
\newblock Long-tailed classification by keeping the good and removing the bad momentum causal effect.
\newblock {\em Advances in neural information processing systems}, 33:1513--1524, 2020.

\bibitem{zhang2021distribution}
Songyang Zhang, Zeming Li, Shipeng Yan, Xuming He, and Jian Sun.
\newblock Distribution alignment: A unified framework for long-tail visual recognition.
\newblock In {\em Proceedings of the IEEE/CVF conference on computer vision and pattern recognition}, pages 2361--2370, 2021.

\bibitem{zhang2022self}
Yifan Zhang, Bryan Hooi, Lanqing Hong, and Jiashi Feng.
\newblock Self-supervised aggregation of diverse experts for test-agnostic long-tailed recognition.
\newblock {\em Advances in Neural Information Processing Systems}, 35:34077--34090, 2022.

\bibitem{wang2023customized}
Fang Wang, Di~Tian, and Mark Carroll.
\newblock Customized deep learning for precipitation bias correction and downscaling.
\newblock {\em Geoscientific Model Development}, 16(2):535--556, 2023.

\bibitem{trebing2021smaatunetprecipitationnowcastingusing}
Kevin Trebing, Tomasz Stanczyk, and Siamak Mehrkanoon.
\newblock Smaat-unet: Precipitation nowcasting using a small attention-unet architecture, 2021.

\bibitem{badrinarayanan2017segnet}
Vijay Badrinarayanan, Alex Kendall, and Roberto Cipolla.
\newblock Segnet: A deep convolutional encoder-decoder architecture for image segmentation.
\newblock {\em IEEE transactions on pattern analysis and machine intelligence}, 39(12):2481--2495, 2017.

\bibitem{li2023moganet}
Siyuan Li, Zedong Wang, Zicheng Liu, Cheng Tan, Haitao Lin, Di~Wu, Zhiyuan Chen, Jiangbin Zheng, and Stan~Z Li.
\newblock Moganet: Multi-order gated aggregation network.
\newblock In {\em The Twelfth International Conference on Learning Representations}, 2023.

\bibitem{andrychowicz2023deep}
Marcin Andrychowicz, Lasse Espeholt, Di~Li, Samier Merchant, Alexander Merose, Fred Zyda, Shreya Agrawal, and Nal Kalchbrenner.
\newblock Deep learning for day forecasts from sparse observations.
\newblock {\em arXiv preprint arXiv:2306.06079}, 2023.

\bibitem{menon2020long}
Aditya~Krishna Menon, Sadeep Jayasumana, Ankit~Singh Rawat, Himanshu Jain, Andreas Veit, and Sanjiv Kumar.
\newblock Long-tail learning via logit adjustment.
\newblock {\em arXiv preprint arXiv:2007.07314}, 2020.

\bibitem{mittermaier2013long}
Marion Mittermaier, Nigel Roberts, and Simon~A Thompson.
\newblock A long-term assessment of precipitation forecast skill using the fractions skill score.
\newblock {\em Meteorological Applications}, 20(2):176--186, 2013.

\bibitem{joyce2004cmorph}
Robert~J Joyce, John~E Janowiak, Phillip~A Arkin, and Pingping Xie.
\newblock Cmorph: A method that produces global precipitation estimates from passive microwave and infrared data at high spatial and temporal resolution.
\newblock {\em Journal of hydrometeorology}, 5(3):487--503, 2004.

\bibitem{feng2024enhancing}
Jin Feng and Yanjie Li.
\newblock Enhancing surface wind speed and temperature prediction using surface-layer emulator and transfer learning.
\newblock {\em Monthly Weather Review}, 152(10):2361--2377, 2024.

\end{thebibliography}

\end{document}